\documentclass[a4paper]{article}
\usepackage[cp866]{inputenc}
\usepackage{graphicx}
\date{}
\author{N.\,V.\,\,Bakhar\thanks{e-mail address: kbakhar@yandex.ru}, 
\,and N.\,V.\,\,Ustinov$\mbox{}^{*,}$\thanks{e-mail address: 
n\_ustinov@mail.ru}
\medskip\\
Faculty of Physics, I.\,\,Kant Russian State University,\\\medskip
A.\,Nevskogo str.\,\,14, Kaliningrad, 236041, Russia\\
$\mbox{}^*$Quantum Field Theory Department, Tomsk State University,\\
36 Lenin Avenue, Tomsk, 634050, Russia
\vspace{-4ex}}
\title{\bf Dynamics of two-component electromagnetic and acoustic extremely 
short pulses}
\date{ }
\begin{document}
\maketitle
\begin{abstract}
\noindent
The distinctive features of passing the two-component extremely short pulses 
through the nonlinear media are discussed.
The equations considered describe the propagation in the two-level anisotropic
medium of the electromagnetic pulses consisting of ordinary and extraordinary
components and an evolution of the transverse-longitudinal acoustic pulses in
a crystal containing the paramagnetic impurities with effective spin $S=1/2$.
The solutions decreasing exponentially and algebraically are studied.
\end{abstract}


\section{Introduction}
In recent years, an important subject of the theoretical studies became the 
coherent phenomena in the anisotropic media.
This is caused by the significant progress in the technologies of growing the
semiconductor crystals and producing the low-dimensional quantum structures.
Unlike to the case of isotropic media, the parity of the stationary states of
the quantum particles located in the anisotropic medium is not well defined.
As a result, the diagonal elements of the matrix of the dipole moment operator
and their difference usually called the permanent dipole moment (PDM) of the
transition may not vanish.

The dynamics of one-component extremely short electromagnetic pulses in the
presence of PDM was considered in works$\mbox{}^{1-5}$.
The stationary pulse solutions decreasing exponentially and algebraically of
the full Maxwell--Bloch equations for the two-level medium were
found$\mbox{}^4$.
It was shown that these solutions possess an asymmetry on the pulse polarity. 
The complete integrability of the corresponding system of the reduced
Maxwell--Bloch equations with the help of the inverse scattering
transformation method$\mbox{}^{6-8}$ was revealed in paper$\mbox{}^2$.
The multisoliton solutions were constructed on fixed background, which was
selected in such a manner that the problem under consideration is reduced to
isotropic case.
The effects of a pump on dynamics of these pulses were discussed
in$\mbox{}^3\!$.
Also, the numerical study of the pulse formation in the medium possessing PDM
was performed$\mbox{}^5$.
An existence of solitary stable bipolar signal with nonzero time area was
discovered.
The asymmetry on the polarity manifested itself there in that the sign of the 
time area is determined by the sign of PDM.

The propagation through optically uniaxial medium of the two-component
electromagnetic pulses consisting of short-wavelength ordinary and
long-wavelength extraordinary components was investigated in$\mbox{}^{9-12}$.
As opposite to the case of one-component pulses mentioned above, an 
application of the slowly varying envelope approximation does not eliminate 
the influence of PDM in this problem.
It was shown that, under conditions of a strong interaction between the
components, the regimes of the pulse propagation may differ from the
self-induced transparency.
Various regimes of the optical transparency were classified with respect to
the pulse velocity and the degree of the resonant medium excitation.
The two-component pulse solutions possess also the PDM-induced asymmetry:
signs of the pulse extraordinary component and PDM are opposite.
It is remarkable that the equations distinguishing from ones studied
in$\mbox{}^{9-12}$ by notations only describe the two-component
(transverse-longitudinal) acoustic pulse propagation in the low-temperature
paramagnetic crystal$\mbox{}^{13,14}$.

To investigate the propagation of the two-component pulses in optically
uniaxial medium, we apply here the approximation of the unidirectional
propagation instead of the slowly varying envelope approximation.
The two-component system of reduced Maxwell--Bloch equations is derived and, as
in the case treated in works$\mbox{}^{9-12}$, its acoustic analogue is found. 
The properties of exponentially and algebraically decreasing pulse solutions
of these equations are discussed.

\section{Two-component system of the Maxwell--Bloch equations for an
anisotropic medium and its acoustic analogue}

Consider the optically uniaxial medium containing the two-level quantum
particles.
Let the medium optical axes coincide with $z$ axes of the Cartesian coordinate
system.
Suppose that the plane electromagnetic pulse propagates through the medium in
the positive direction of $y$ axes.
Then, ordinary $E_o$ and extraordinary $E_e$ components of the electric field
of the pulse are parallel to $x$ and $z$ axes, respectively.
The Maxwell equations yield in the case we study the next system for the
electric field components:
\begin{equation}
\displaystyle\frac{\partial^2 E_o}{\partial y^2}-
\frac{n^2_o}{c^2}\frac{\partial^2 E_o}{\partial t^2}
=\frac{4\pi}{c^2}\frac{\partial^2 P_o}{\partial t^2},
\label{1}
\end{equation}
\begin{equation}
\displaystyle\frac{\partial^2 E_e}{\partial y^2}-
\frac{n^2_e}{c^2}\frac{\partial^2 E_e}{\partial t^2}
=\frac{4\pi}{c^2}\frac{\partial^2 P_e}{\partial t^2},
\label{2}
\end{equation}
where $P_o$ and $P_e$ are the ordinary and extraordinary components of the
polarization, which is induced by the two-level particles; $n_o$ and $n_e$ are
ordinary and extraordinary refractive indices, respectively; $c$ is the speed
of light in free space.

To describe the evolution of the state of the quantum particle, we make use
the von~Neumann equation on density matrix $\hat\rho$:
\begin{equation}
i\hbar\frac{\partial\hat\rho}{\partial t}=[\hat H,\hat\rho].
\label{3}
\end{equation}
Here Hamiltonian $H$ of the two-level particle is defined as given
\begin{equation}
H=\mbox{diag}(0,\hbar\omega_0)-\hat d_xE_o-\hat d_zE_e,
\label{4}
\end{equation}
where $\omega_0$ is the resonant frequency of quantum transition
$\mid\!1\rangle\!\!\to\mid\!2\rangle$; $\hbar$ is the Plank's constant;
$\hat d_x$ and $\hat d_z$ are the matrices of the projection of the dipole
moment operator on $x$ and $z$ axes, respectively.

In the geometry chosen, the expressions for the polarization components are
\begin{equation}
P_o=N\,\mbox{Tr}\,(\hat\rho\hat d_x),
\label{5}
\end{equation}
\begin{equation}
P_e=N\,\mbox{Tr}\,(\hat\rho\hat d_z),
\label{6}
\end{equation}
where $N$ is the concentration of the quantum particles.
Since the medium is axially symmetric, the matrices of the dipole moment have
the following form:
\begin{equation}
\hat d_x=\left(
\begin{array}{cc}
0&d_{12}\\
d_{12}&0
\end{array}
\right)\!,
\label{7}
\end{equation}
\begin{equation}
\hat d_z=\left(
\begin{array}{cc}
D_{11}&0\\
0&D_{22}
\end{array}
\right)\!.
\label{8}
\end{equation}
Then Hamiltonian (\ref{4}) reads as
\begin{equation}
H=\left(
\begin{array}{cc}
-D_{11}E_e&-d_{12}E_o\\
-d_{12}E_o&\hbar\omega_0-D_{22}E_e
\end{array}
\right)\!,
\label{9}
\end{equation}
and equation (\ref{3}) gives the next system for the density matrix elements:
\begin{equation}
\frac{\partial\rho_{11}}{\partial t}=-\frac{\partial\rho_{22}}{\partial t}=
i\frac{d_{12}}{\hbar}E_o(\rho_{21}-\rho_{12}),
\label{10a}
\end{equation}
\begin{equation}
\frac{\partial\rho_{12}}{\partial t}=i\left(\omega_0+\frac{D}{\hbar}E_e\right)
\rho_{12}+i\frac{d_{12}}{\hbar}E_o(\rho_{22}-\rho_{11}),
\label{11a}
\end{equation}
where $D=D_{11}-D_{22}$ is PDM of the transition.
Formulas (\ref{1}), (\ref{2}) and (\ref{5})--(\ref{8}) imply
\begin{equation}
\displaystyle\frac{\partial^2E_o}{\partial y^2}-
\frac{n^2_o}{c^2}\frac{\partial^2E_o}{\partial t^2}=\frac{4\pi N}{c^2}d_{12}
\frac{\partial^2}{\partial t^2}(\rho_{12}+\rho_{21}),
\label{10}
\end{equation}
\begin{equation}
\displaystyle\frac{\partial^2E_e}{\partial y^2}-
\frac{n^2_e}{c^2}\frac{\partial^2E_e}{\partial t^2}=
\frac{4\pi N}{c^2}\frac{\partial^2}{\partial t^2}
(D_{11}\rho_{11}+D_{22}\rho_{22}).
\label{11}
\end{equation}

Let us suppose that concentration of the quantum particles is small
($\pi d_{12}^2N/(\hbar\omega_0)\!\ll\!1$).
The order of derivatives in the equations for the electric field components
can be reduced in this case.
Indeed, by applying the approximation of the unidirectional
propagation$\mbox{}^{15}$ to equations (\ref{10}), (\ref{11}) and excluding
the derivatives of the elements of the density matrix with the help of
formulas (\ref{10a}) and (\ref{11a}), we obtain:
\begin{equation}
\frac{\partial E_o}{\partial y}+\frac{n_o}{c}\frac{\partial
E_o}{\partial t}=-i\frac{2\pi N}{n_o
c}d_{12}\left(\omega_0+\frac{D}{\hbar}E_e\right)
(\rho_{12}-\rho_{21}),
\label{12}
\end{equation}
\begin{equation}
\frac{\partial E_e}{\partial y}+\frac{n_e}{c}\frac{\partial
E_e}{\partial t}=i\frac{2\pi N E_o}{n_e c
\hbar}d_{21}D(\rho_{12}-\rho_{21}).
\label{13}
\end{equation}

It is convenient for the subsequent consideration to rewrite equations
(\ref{10a}), (\ref{11a}) and (\ref{12}), (\ref{13}) in dimensionless form
\begin{equation}
\frac{\partial u}{\partial\xi}=i(1+v)(\sigma^*-\sigma),
\label{14}
\end{equation}
\begin{equation}
\frac{\partial v}{\partial\xi}+\delta_n\frac{\partial v}{\partial\tau}=
ik^2\frac{n_o}{n_e}\,u(\sigma-\sigma^*),
\label{15}
\end{equation}
\begin{equation}
\frac{\partial\sigma_3}{\partial\tau}=iu(\sigma-\sigma^*),
\label{16}
\end{equation}
\begin{equation}
\frac{\partial\sigma}{\partial\tau}=i(1+v)\sigma+2iu\sigma_3,
\label{17}
\end{equation}
where
$$
\tau=\omega_0\left(t-\frac{n_o}{c}y\right),\quad
\xi=\frac{2\pi Nd_{12}^2}{\hbar n_oc}y,
$$
\begin{equation}
u=\frac{d_{12}}{\hbar\omega_0}E_o,\quad
v=\frac{D}{\hbar\omega_0}E_e,
\label{19}
\end{equation}
$$
\sigma_3=\frac{\rho_{22}-\rho_{11}}2,\quad
\sigma=\rho_{12},
$$
$$
k=\frac{D}{d_{12}},\quad
\delta_n=\frac{n_o\hbar\omega_0}{2\pi N d_{1 2}^2}(n_e-n_o).
$$

Obviously, system (\ref{14})--(\ref{17}) coincides with the reduced 
Maxwell--Bloch equations for isotropic medium if $k=v=0$. 
One can see that the electric field components fulfill different functions 
here.
The ordinary component causes the quantum transitions, while the extraordinary 
one shifts its frequency. 
Also, the same functions were executed by these components in 
works$\mbox{}^{9-12}$, where the system, which follows equations 
(\ref{14})--(\ref{17}) in the slowly varying envelope approximation, was 
considered.

It is remarkable that the system of equations equivalent to 
(\ref{14})--(\ref{17}) appeared in the acoustics. 
Namely, the propagation of the transverse-longitudinal acoustic pulses in a 
crystal containing paramagnetic impurities with effective spin $S=1/2$ in a 
direction parallel to the external magnetic field is described by next 
equations (see$\mbox{}^{16,17}$): 
\begin{equation}
\frac{\partial E}{\partial\chi}=-bUS_y,
\label{22}
\end{equation}
\begin{equation}
\frac{\partial U}{\partial\chi}+\delta_v\frac{\partial U}{\partial\eta}=
bES_y,
\label{23}
\end{equation}
\begin{equation}
\frac{\partial S_y}{\partial\eta}=bUS_x-ES_z,
\label{24}
\end{equation}
\begin{equation}
\frac{\partial S_x}{\partial\eta}=-bUS_y,
\label{25}
\end{equation}
\begin{equation}
\frac{\partial S_z}{\partial\eta}=ES_y,
\label{26}
\end{equation}
where
$$
\delta_v=(v_1-v_2)\frac{v_1n_0g^2}{\hbar\omega_Bnf_3^2}.
$$
$E$ and $U$ are dimensionless variables defining transverse and longitudinal 
components of the acoustic pulse, respectively; $S_x$, $S_y$ and $S_z$ are 
expressed through the elements of the density matrix of the paramagnetic 
impurities; $\chi$ and $\eta$ are dimensionless spatial and retarded time 
variables; constant $b$ is defined through the coupling constants of the 
spin-phonon interaction; $v_1$ and $v_2$ are linear velocities of the 
transverse and longitudinal acoustic waves, respectively; $\omega_B$ is the 
frequency of the Zeeman splitting of the Kramers's doublets; $n$ is the 
concentration of the paramagnetic impurities; $n_0$ is the mean crystal 
density; $g$ is the component of the Lande tensor; $f_3$ is the coupling 
constant of the spin-phonon interaction.

The variables of both systems are connected by relations:
\begin{equation}
E=ku,\quad U=\sqrt{\frac{n_e}{n_o}}\,(1+v),
\label{er1}
\end{equation}
\begin{equation}
S_x=-k(\sigma+\sigma^*),\quad S_y=ik(\sigma-\sigma^*),\quad
S_z=2k\sigma_3,
\label{er2}
\end{equation}
\begin{equation}
\chi=\frac{2}{k}\xi,\quad\eta=\frac{2}{k}\tau,
\label{er3}
\end{equation}
\begin{equation}
b=\sqrt{\frac{n_o}{n_e}}\,\frac{k}{2},\quad\delta_v=\delta_n.
\label{er4}
\end{equation}

Complete integrability of equations (\ref{22})--(\ref{26}) in the frameworks 
of the inverse scattering transformation method$\mbox{}^{6-8}$ in the case, 
when the linear velocities of transverse and longitudinal acoustic waves are 
equal, was established in paper$\mbox{}^{17}$. 
Consequently, system (\ref{14})--(\ref{17}) is also integrable if 
$\delta_n=0$.

\section{Simplest solutions}

Let us begin with system (\ref{14})--(\ref{17}). 
At first, we suppose that the refractive indices of the medium are equal: 
$n_e=n_o$. 
As it was noted at the end of previous section, system we deal with is 
integrable by inverse scattering transformation method in this case, and its 
multisoliton solutions can be obtained by means of the algebraic methods of 
the soliton theory. 
The Darboux transformation$\mbox{}^{18}$ technique can be used for 
constructing these solutions, for instance. 
Since the one-soliton solution is stationary, it can be found by direct 
integration of equations (\ref{14})--(\ref{17}). 
After straightforward calculations, one reveals that there can exist two 
families of the one-soliton solutions. 
These families are distinguished by the domains on the spectral parameter 
plane, in which the points of the discrete spectrum of the solution spectral 
data lie. Solution of the first family has next form:
\begin{equation}
u=\frac{\sqrt{A}\,\sinh\!\theta}{A\sinh^2\!\theta+K},
\label{27}
\end{equation}
\begin{equation}
v=-\frac{2K}{A\sinh^2\!\theta+K},
\label{28}
\end{equation}
\begin{equation}
\sigma=-\frac{2T\sqrt{A}\,\sigma_0(T\sinh\!\theta+i\cosh\!\theta)}
{(1+T^2)(A\sinh^2\!\theta+K)},
\label{29}
\end{equation}
\begin{equation}
\sigma_3=\sigma_0\left(1-\frac{2T^2}{(1+T^2)(A\sinh^2\!\theta+K)}\right),
\label{30}
\end{equation}
where
$$
K=\frac{k^2}{4},\quad A=K(1+T^2)-T^2,
$$
$$
\theta=\frac{\tau}{T}+\frac{4\sigma_0 T}{1+T^2}\xi,
$$
$\sigma_0$ is an initial population of the quantum level ($|\sigma_0|\le1/2$).
Free parameter of this solution is real constant $T$.
It is assumed here that parameter $T$ takes the values such that condition
$A>0$ holds.
This condition is true with arbitrary $T$, if $K\ge1$.
For $K<1$, parameter $T$ has to satisfy constraint
$$
T^2\le\frac{K}{1-K}.
$$

In the case $A<0$, the one-soliton solution is written as follows
\begin{equation}
u=\frac{\sqrt{-A}\,\cosh\!\theta}{A\cosh^2\!\theta-K},
\label{31}
\end{equation}
\begin{equation}
v=\frac{2K}{A\cosh^2\!\theta-K},
\label{32}
\end{equation}
\begin{equation}
\sigma=-\frac{2T\sqrt{-A}\,\sigma_0(T\cosh\!\theta+i\sinh\!\theta)}
{(1+T^2)(A\cosh^2\!\theta-K)},
\label{33}
\end{equation}
\begin{equation}
\sigma_3=\sigma_0\left(1+\frac{2T^2}{(1+T^2)(A\cosh^2\!\theta-K)}\right).
\label{34}
\end{equation}
This solution belongs to the second family of the one-soliton solutions.
It can exist only if $K<1$.
Also, the parameter $T$ value is such that
$$
T^2>\frac{K}{1-K}.
$$
The well-known one-soliton solution of the reduced Maxwell--Bloch 
equations$\mbox{}^{15}$ is obtained from formulas (\ref{31})--(\ref{34}) if we 
put $k=0$.

It is seen from equations (\ref{19}), (\ref{28}) and (\ref{32}) that there
exists asymmetry on the polarity of the extraordinary component of the pulse
electric field: signs of $E_e$ and PDM are opposite.
Similar asymmetry was found in$\mbox{}^{9-12}$ for two-component pulses, in
which the ordinary component has slowly varying envelope.
The number of peaks of variable $u$ of the second family solution depends on
parameter $T$.
If
$$
T^2\le\frac{2K}{1-K}
$$
it has two peaks, while at
$$
T^2>\frac{2K}{1-K}
$$
only single peak exists.
One can show that a degree of the medium excitation grows with increase of
$|T|$ for the first one-soliton solution family and, on the contrary, with
decrease of $|T|$ for second one.
The full inversion of the quantum level population can take place only if
$K\le1$.
The curves of variables $u$ and $\sigma_3$ of the first and second families of
one-soliton solution are presented on Figs 1, 2 and Fig. 3, respectively. 
The distance between the peaks of $u$ depends on parameter $T$ and can take an 
arbitrary value if $K<1$ (see curves on Figs~2a and 3a). 
If $K>1$, there exists an upper limit of this distance that corresponds to 
Fig.~6 below (compare with Fig.~1a). 
The strongest excitation of the quantum particles occurs in the center of the 
pulses in all cases. 
\begin{figure}[ht]
\centering
\includegraphics[width=3.34in]{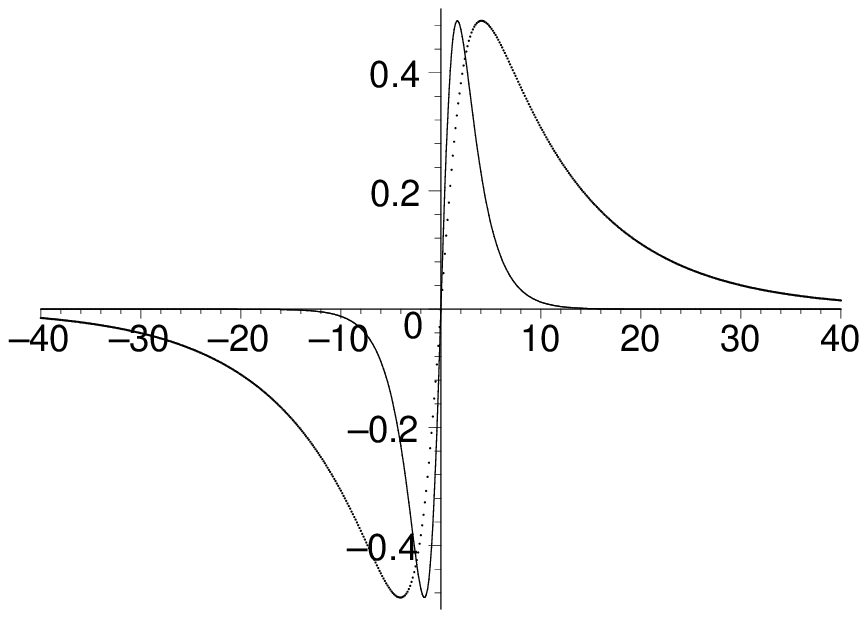}
\includegraphics[width=3.34in]{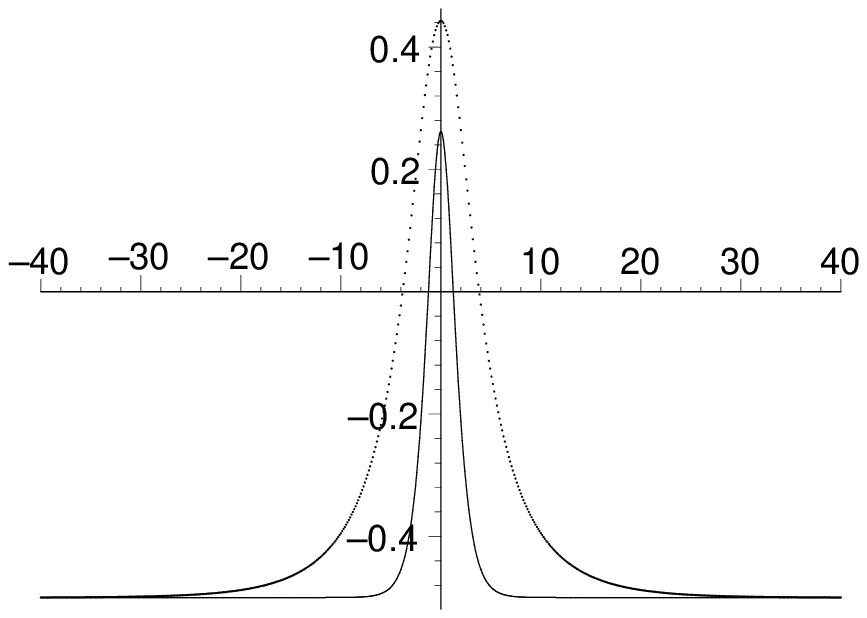}
\begin{picture}(0,0)
\put(-135,170){$u$}
\put(-35,108){$\tau$}
\put(130,170){$\sigma_3$}
\put(205,98){$\tau$}
\put(-215,170){\small\bf a}
\put(30,170){\small\bf b}
\end{picture}
\vskip-0.5cm \centerline{\parbox{15.0cm}{\small\bf Fig.\,1. \rm Profiles of 
$u$ and $\sigma_3$ of the first one-soliton family with $K=1.05$, 
$\sigma_0=-0.5$, $T=2$ (solid curves) and $T=10$ (dotted lines).}}
\end{figure}
\begin{figure}[ht]
\centering
\includegraphics[width=3.34in]{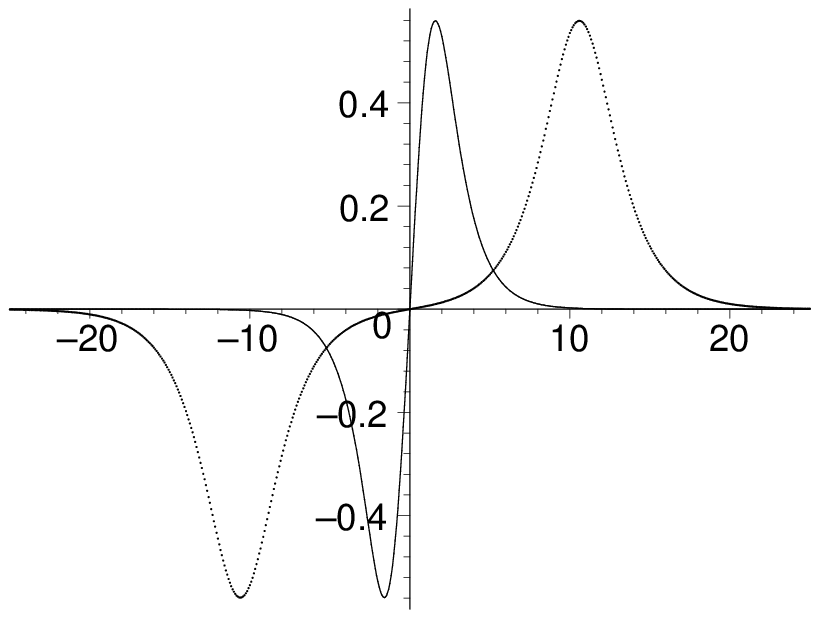}
\includegraphics[width=3.34in]{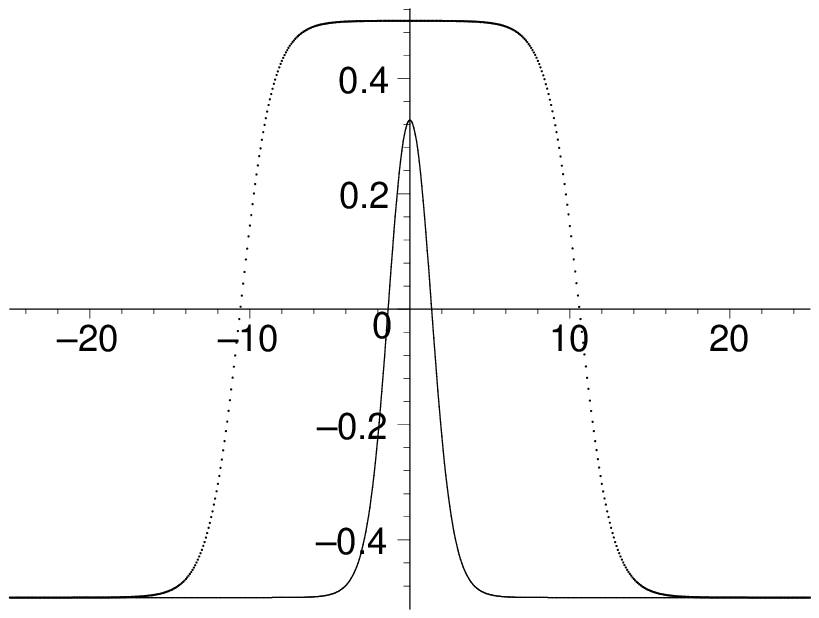}
\begin{picture}(0,0)
\put(-135,170){$u$}
\put(-35,108){$\tau$}
\put(130,178){$\sigma_3$}
\put(205,108){$\tau$}
\put(-215,170){\small\bf a}
\put(30,170){\small\bf b}
\end{picture}
\vskip-0.5cm \centerline{\parbox{15.0cm}{\small\bf Fig.\,2. \rm Profiles of 
$u$ and $\sigma_3$ of the first one-soliton family with $K=0.8$, 
$\sigma_0=-0.5$, $T=1.4$ (solid curves) and $T=1.9999$ (dotted curves).}}
\end{figure}
\begin{figure}[ht]
\centering
\includegraphics[width=3.34in]{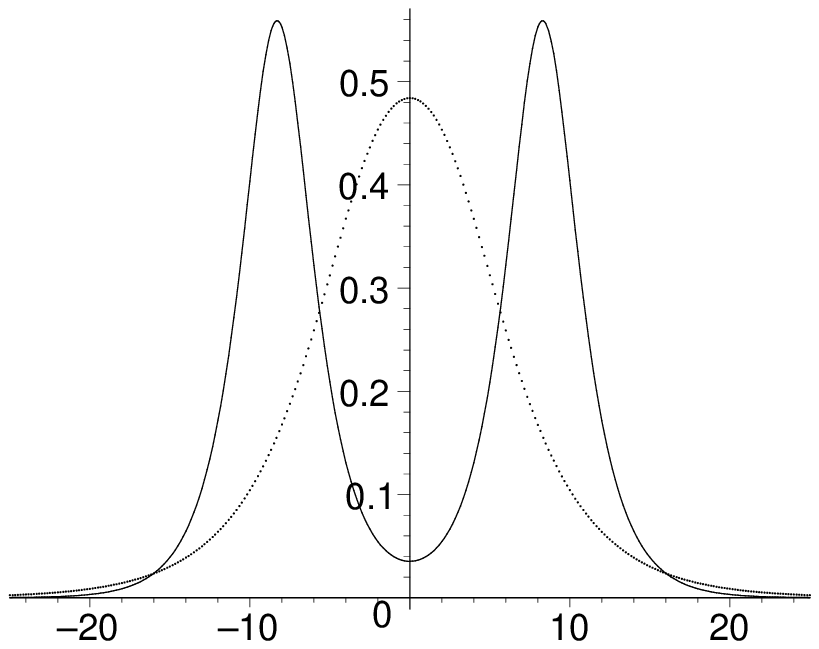}
\includegraphics[width=3.34in]{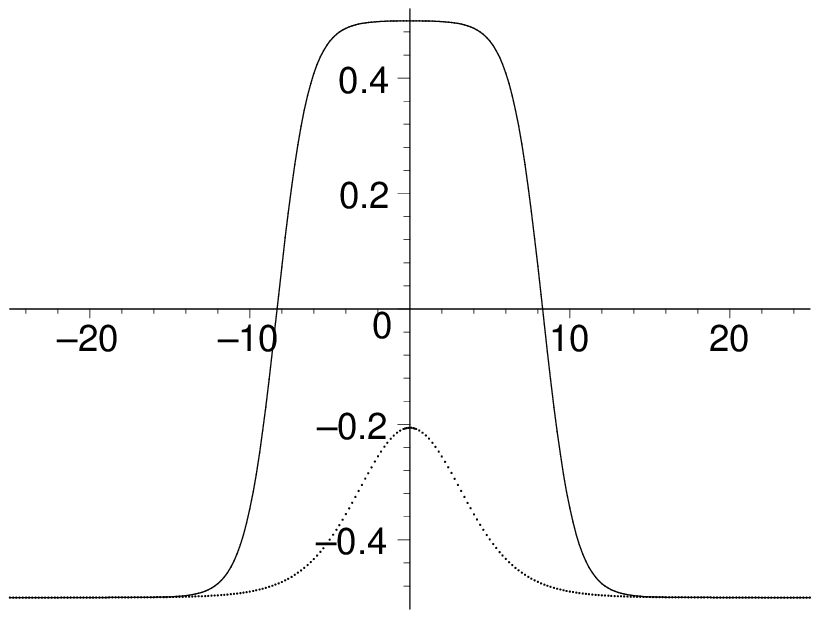}
\begin{picture}(0,0)
\put(-135,170){$u$}
\put(-35,38){$\tau$}
\put(130,178){$\sigma_3$}
\put(205,108){$\tau$}
\put(-215,170){\small\bf a}
\put(30,170){\small\bf b}
\end{picture}
\vskip-0.5cm \centerline{\parbox{15.0cm}{\small\bf Fig.\,3. \rm Profiles of 
$u$ and $\sigma_3$ of the second one-soliton family with $K=0.8$, 
$\sigma_0=-0.5$, $T=2.001$ (solid curves) and $T=4$ (dotted lines).}}
\end{figure}

Consider different limiting cases of the one-soliton solutions.
If $K<1$, we can tend $A$ to zero.
Then, after appropriate shift alone $\tau$ (or $\xi$) axes, we obtain from
formulas (\ref{27})--(\ref{30}) or (\ref{31})--(\ref{34}) following
expressions:
\begin{equation}
u=\frac{T\exp\theta}{T^2+K\exp2\theta},
\label{35}
\end{equation}
\begin{equation}
v=-\frac{2K\exp2\theta}{T^2+K\exp2\theta},
\label{36}
\end{equation}
\begin{equation}
\sigma=-\frac{2T^2\sigma_0(T-i)\exp\theta}{(1+T^2)(T^2+K\exp2\theta)},
\label{37}
\end{equation}
\begin{equation}
\sigma_3=\sigma_0\left(1-\frac{2T^2\exp2\theta}
{(1+T^2)(T^2+K\exp2\theta)}\right).
\label{38}
\end{equation}
It is necessary to put $T=k/\sqrt{4-k^2}$ in this solution in accordance with
the limiting procedure.
\begin{figure}[ht]
\vskip0.0cm
\centering
\includegraphics[width=3.5in]{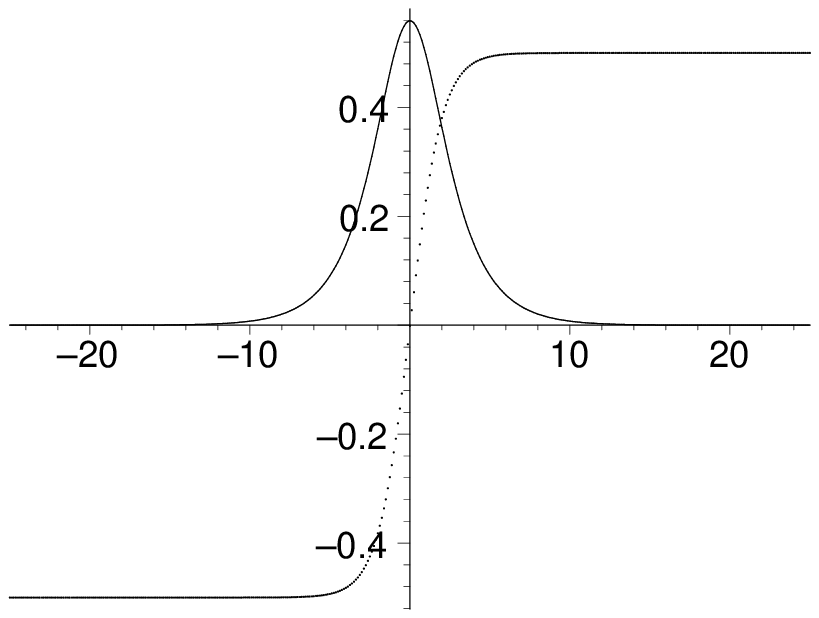}
\begin{picture}(0,0)
\put(-121,170){$u$, $\sigma_3$}
\put(-40,96){$\tau$}
\end{picture}
\vskip-0.5cm \centerline{\parbox{10.0cm}{\small\bf Fig.\,4. \rm Profiles of 
$u$ (solid curve) and $\sigma_3$ (dotted curve) in the case $A=0$ with 
$K=0.8$, $T=2.0$, $\sigma_0=-0.5$.}}
\end{figure}
\begin{figure}[ht]
\vskip0.0cm
\centering
\includegraphics[width=3.5in]{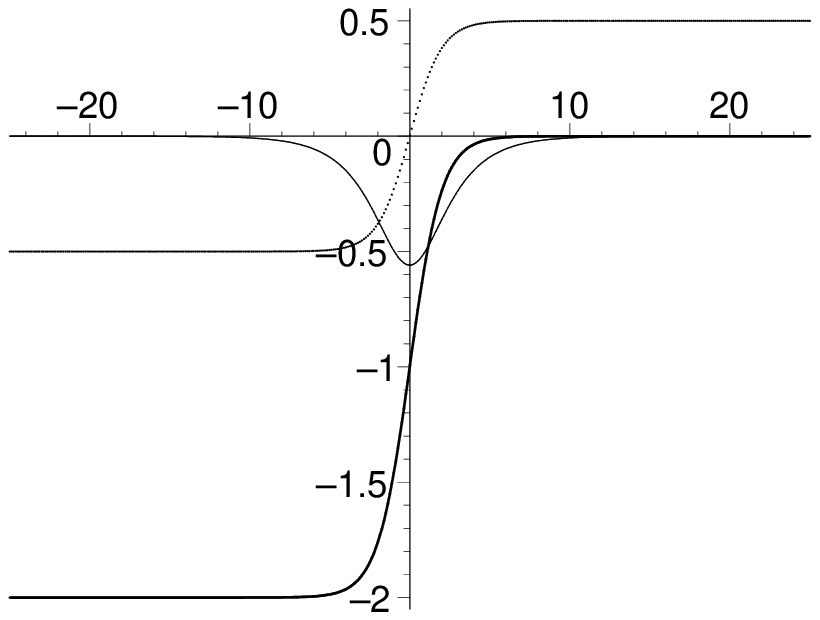}
\begin{picture}(0,0)
\put(-121,173){$u$, $v$, $\sigma_3$}
\put(-36,130){$\tau$}
\end{picture}
\vskip-0.5cm \centerline{\parbox{10.0cm}{\small\bf Fig.\,5. \rm Profiles of 
$u$ (solid curve), $v$ (thick curve) and $\sigma_3$ (dotted line) in the case 
$A=0$ with $K=0.8$, $T=-2.0$, $\sigma_0=0.5$.}}
\end{figure}
If $|T|\to\infty$, then formulas (\ref{27})--(\ref{30}) lead to algebraic
solution:
\begin{equation}
u=\frac{\sqrt{K-1}\,(\tau+4\sigma_0\xi)}{(K-1)(\tau+4\sigma_0\xi)^2+K},
\label{39}
\end{equation}
\begin{equation}
v=-\frac{2K}{(K-1)(\tau+4\sigma_0\xi)^2+K},
\label{40}
\end{equation}
\begin{equation}
\sigma=-\frac{2\sigma_0\sqrt{K-1}(\tau+4\sigma_0\xi+i)}
{(K-1)(\tau+4\sigma_0\xi)^2+K},
\label{41}
\end{equation}
\begin{equation}
\sigma_3=\sigma_0\left(1-\frac{2}{(K-1)(\tau+4\sigma_0\xi)^2+K}\right).
\label{42}
\end{equation}
This limit is allowed in the case $K>1$ only.
\begin{figure}[ht]
\vskip0.0cm
\centering
\includegraphics[width=3.5in]{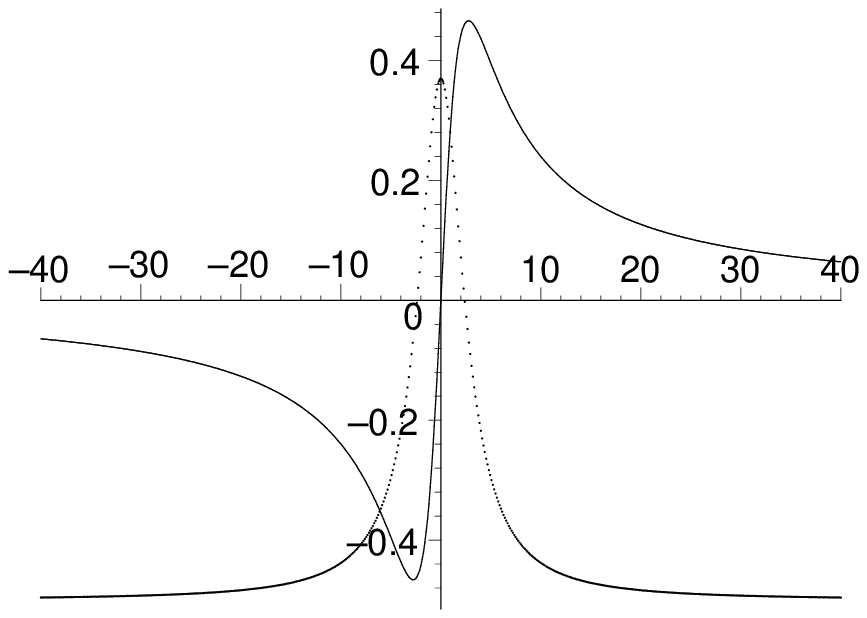}
\begin{picture}(0,0)
\put(-157,170){$u$, $\sigma_3$}
\put(-40,88){$\tau$}
\end{picture}
\vskip-0.5cm \centerline{\parbox{10.0cm}{\small\bf Fig.\,6. \rm Profiles of 
$u$ (solid curve) and $\sigma_3$ (dotted line) of the algebraic solution with 
$K=1.15$, $\sigma_0=-0.5$.}}
\end{figure}

Dependence of variables $u$ and $\sigma_3$ of solutions (\ref{35})--(\ref{38}) 
and (\ref{39})--(\ref{42}) on $\tau$ are presented in Figs~4,~5 and 6, 
respectively. 
Comparing with curves on Figs~2a and 3a, we can say for the variable $u$ case 
that the first limiting procedure ($A\to0$) corresponds to moving one of its 
peaks into infinity. 
The pulse, whose velocity exceeds the light velocity, is plotted in Fig.~5. 
Note that variable $v$ tends to $-2$ in the limit $\tau\to-\infty$. 
For this reason, the energy levels trade their places owing to the Stark 
effect. 

Let us now consider the case of unequal refractive indices $(n_e\ne n_o)$.
If we define coefficient $K$ as
\begin{equation}
K=\frac{n_ok^2}{4n_e}\left(1+\frac{1+T^2}{4\sigma_0
T^2}\delta_n\right)^{-1}
\label{43}
\end{equation}
in the formulas (\ref{27})--(\ref{30}) and (\ref{31})--(\ref{34}), then they 
give solutions of system (\ref{14})--(\ref{17}) with $\delta_n\ne0$.
Since coefficient $K$ characterizing the anisotropy of the medium depends now 
on the parameter of the pulse, we can say that the anisotropy becomes 
effective. 
It is seen that this coefficient is unbounded if
$$
T^2=-\frac{1}{1+4\sigma_0/\delta_n}
$$
and it can change the sign.
Similar results were obtained in works$\mbox{}^{9-12}$, where the 
approximation of the slowly varying envelope was applied for the description 
of the dynamics of the two-component pulses. 

Obviously, the limiting cases studied above can be also modified to satisfy
equations (\ref{14})--(\ref{17}) with $n_e\ne n_o$.
In particular, one has to put
$$
K=\frac{n_ok^2}4{n_e}\left(1+\frac{\delta_n}{4\sigma_0}\right)^{-1},
$$
in formulas (\ref{39})--(\ref{42}).
Finally, coefficient $K$ should be determined by relation (\ref{43}) in
equations (\ref{35})--(\ref{38}), while parameter $T$ should be equal to
\begin{equation}
T=\pm\sqrt{\frac{\symbol{26}}{4-\symbol{26}}},
\label{45}
\end{equation}
where
$$
\symbol{26}=\frac{n_o}{n_e}k^2-\frac{\delta_n}{\sigma_0}.
$$

The solutions of system (\ref{22})--(\ref{26}) can be found using its 
equivalence to system (\ref{14})--(\ref{17}) (see relations 
\mbox{(\ref{er1})--(\ref{er4})).} 
In particular, formulas (\ref{27})--(\ref{30}) and (\ref{31})--(\ref{34}) 
give two families of the pulse solutions of the acoustic system:
$$
E=\frac{2\,\sqrt{B}\,\sinh\!\Theta}{B\sinh^2\!\Theta+1},
$$
$$
U=1-\frac{2}{B\sinh^2\!\Theta+1},
$$
$$
S_x=\frac{2b\,\sqrt{B}\,{\hat T}^2S_0\sinh\!\Theta}
{(1+b^2{\hat T}^2)(B\sinh^2\!\Theta+1)},
$$
$$
S_y=\frac{2\sqrt{B}\,\hat TS_0\cosh\!\Theta}{(1+b^2{\hat T}^2)
(B\sinh^2\!\Theta+1)},
$$
$$
S_z=S_0\left(1-\frac{2{\hat T}^2}{(1+b^2{\hat T}^2)
(B\sinh^2\!\Theta+1)}\right),
$$
and
$$
E=\frac{2\,\sqrt{-B}\,\cosh\!\Theta}{B\cosh^2\!\Theta-1},
$$
$$
U=1+\frac{2}{B\cosh^2\!\Theta-1},
$$
$$
S_x=\frac{2b\,\sqrt{-B}\,{\hat T}^2S_0\cosh\!\Theta}
{(1+b^2{\hat T}^2)(B\cosh^2\!\Theta-1)},
$$
$$
S_y=\frac{2\sqrt{-B}\,\hat TS_0\sinh\!\Theta}{(1+b^2{\hat T}^2)
(B\cosh^2\!\Theta-1)},
$$
$$
S_z=S_0\left(1+\frac{2{\hat T}^2}{(1+b^2{\hat T}^2)
(B\cosh^2\!\Theta-1)}\right),
$$
where 
$$
B=1+(b^2-1){\hat T^2},\quad 
\Theta=\frac{\eta}{\hat T}+\frac{b\hat TS_0}{1+b^2{\hat T}^2}\chi,
$$
$S_0$ is an initial population of the quantum level, real constant $\hat T$ is 
free parameter of these solutions. 
Coefficient $B$ has to be positive for the former solution and negative for 
latter one. 
For the sake of simplicity, we suppose here that $\delta_v=0$. 
The main properties of the pulse solutions presented are, obviously, the same 
as for the system (\ref{14})--(\ref{17}) solutions. 
The appropriate modification of $U$ and $b$ in these formulas allows one to 
obtain the solutions of system (\ref{22})--(\ref{26}) with $\delta_v\ne0$.

\section{Conclusion}

The two-component system of the reduced Maxwell--Bloch equations for 
anisotropic medium and its acoustic analogue have been considered in this 
report. 
As it was done in works$\mbox{}^{11,12}$, the linear velocities of both the
components are not supposed to be equal. 
Two families of the one-parameter extremely short pulse solutions and their 
limiting cases, including one that leads to algebraically decreasing 
two-component pulse, are studied. 
The ordinary component of the pulse solutions found can have one peak like in 
the isotropic case or, also, two peaks, whose signs can be identical or 
opposite. 
If PDM is small ($|D/2d|<1$), then the distance between the peaks can be 
unbounded. 
Nevertheless, the strongest degree of the medium excitation is always achieved 
in the center of the pulses. 
The asymmetry on the polarity of the signal takes place here similarly to that 
in the cases of one-component pulses$\mbox{}^{4,5}$ and two-component pulses, 
in which the ordinary component has higher-frequency filling$\mbox{}^{9-12}$. 
Namely, the signs of the extraordinary component and PDM are opposite.

\section{Acknowledgment}

This work was partially supported by the RBRF grant 05--02--16422\,a.

\end{document}